\documentclass[a4paper]{jpconf}
\usepackage{graphicx}
\usepackage{soul}
\usepackage{colortbl}
\usepackage{multirow}
\usepackage{amsmath,amssymb}
\usepackage{hyperref}% makes the references hyperlinked
\begin{document}
\title{A calculation of the Weyl anomaly for 6D Conformal Higher Spins}

\author{R Aros$^1$, F Bugini$^2$ and D E Diaz$^3$}

\address{$^1$ Departamento de Ciencias F\'isicas, Universidad Andres Bello,
Sazi\'e 2212, Piso 7, Santiago, Chile}
\address{$^2$ Departamento de Matem\'atica y F\'isica Aplicadas,
Universidad Cat\'olica de la Sant\'isima Concepci\'on,
Alonso de Ribera 2850, Concepci\'on, Chile}
\address{$^3$ Departamento de Ciencias F\'isicas, Universidad Andres Bello,
Autopista Concepci\'on-Talcahuano 7100, Talcahuano, Chile} 

\ead{raros@unab.cl, fbugini@ucsc.cl, danilodiaz@unab.cl}

\begin{abstract}
In this work we continue the study of the one-loop partition function for higher derivative conformal higher spin (CHS) fields in six dimensions and its holographic counterpart given by massless higher spin Fronsdal fields in seven dimensions.

In going beyond the conformal class of the boundary round 6-sphere, we start by considering a Ricci-flat, but not conformally flat, boundary and the corresponding Poincar\'e-Einstein space-filling metric. Here we are able to match the UV logarithmic divergence of the boundary with the IR logarithmic divergence of the bulk, very much like in the known 4D/5D setting, under the assumptions of factorization of the higher derivative CHS kinetic operator and  WKB-exactness of the heat kernel of the dual bulk field. A key technical ingredient in this construction is the determination of the fourth heat kernel coefficient $b_6$ for Lichnerowicz Laplacians on both 6D and 7D Einstein manifolds. These results allow to obtain, in addition to the already known type-A Weyl anomaly, two of the three independent type-B anomaly coefficients in terms of the third, say $c_3$ for instance. 

In order to gain access to $c_3$, and thus determine the four central charges independently, we further consider a generic non Ricci-flat Einstein boundary. However, in this case we find a mismatch between boundary and bulk computations for spins higher than two. We close by discussing the nature of this discrepancy and perspectives for a possible amendment.
\end{abstract}

\section{Introduction}

The AdS/CFT correspondence \cite{Maldacena:1997re,Witten:1998qj,Gubser:1998bc} has evolved over the years into a central subject of interest in theoretical physics. While many of the succeeding efforts have been devoted to applications in condensed matter, nuclear and particle physics, it seems fair to say that quantum gravity aspects of the duality are still largely unexplored. The most studied regime so far considers tree level bulk gravity and leading $1/N$ on the boundary.
Genuine quantum gravity effects come out in this regime as a perturbative expansion in terms of loop Witten diagrams, which have proven quite difficult to compute. Nonetheless, there has been recent progress in computing one-loop Witten diagrams (see e.g.~\cite{Giombi:2017hpr,Ponomarev:2019ltz,Carmi:2019ocp}). 

There is yet another instance that probes one-loop quantum effects on both bulk and boundary theories in terms of partition functions. It consists of a class of bulk-boundary fluctuation determinant relation of the type \cite{Gubser:2002zh,Gubser:2002vv}, 
    \begin{equation}\label{holo}
\frac{Z^{^{1-loop,-}}_{_{bulk}}}{Z^{^{1-loop,+}}_{_{bulk}}}\,=\,Z^{^{1-loop}}_{_{bndry}}~.
\end{equation}
The bulk side contemplates the ratio of the functional determinants of the kinetic operator of the bulk field computed with standard and alternate boundary conditions. The boundary side involves the functional determinant of the kinetic operator of the induced field. This kind of {\it holographic formula}  was obtained via a rather circuitous route within AdS/CFT correspondence, in connection with a class of RG flows triggered by double-trace deformations of the CFT~\cite{Gubser:2002zh,Gubser:2002vv,Hartman:2006dy}.
Full match for a massive scalar in Euclidean AdS bulk was first shown in~\cite{Diaz:2007an}, and by now there are plenty of extensions: an incomplete list includes fields with nonzero spin (Dirac, MHS, etc.) and quotients of AdS space (thermal AdS, BTZ, singular AdS, etc.)~\cite{Diaz:2008hy,Bugini:2018def}.

The aim in the present work is to proceed with the analysis of the holographic formula for boundary conformal higher spin fields in six dimensions and to extend the dictionary already established in the conformally flat situation~\cite{Tseytlin:2013jya}. In particular, we explore Ricci-flat but non-conformally flat boundaries as was done in 4D \cite{Acevedo:2017vkk} for CHS and also in 6D for GJMS~\cite{Bugini:2018def} and for the Weyl graviton\cite{Aros_2019JHEP}. In doing so, a crucial technical breakthrough is achieved by the computation of the heat coefficient $b_6$ for Lichnerowicz Laplacians on an Einstein manifold background. We find out a remarkable general combinatorial structure of the coefficient valid in any dimension. More specifically, the 6D results perfectly agree with those obtained via the group theory method in \cite{Liu:2017ruz} for totally symmetric rank-s tensors in 6D. 
We then examine the consequences of the extension of the above mapping to a generic Einstein boundary manifold and discuss the nature of the mismatch we find as well as suggestions for possible amendments.  
\section{Preliminaries: conformal flatness}

In general, the extension of conformal higher spins to curved backgrounds can be quite challenging. However, on conformally flat Einstein backgrounds (e.g.~(A)dS, $S$, $H$) the corresponding Weyl-covariant 2s+2-derivative kinetic operator factorizes into products of standard $2^{\text{nd}}$-derivative operators. 
For the 6D sphere this was explicitly exploited in \cite{Tseytlin:2013fca} to compute the UV-log divergence of the one-loop partition function, and thus the type-A anomaly coefficient, using $\zeta$-function regularization.
According to the holographic formula, the counterpart in the bulk is given by the following ratio of functional determinants on AdS$_7$  
\begin{equation}\label{ConformallyFlatZZZ}
Z^{^{1-loop}}_{_{bulk}}=\left[\frac{\det_{\bot\top}\left\{\hat{\Delta}_L^{(s)}+2(s-1)(s+4)\right\}}
{\det_{\bot\top}\left\{\hat{\Delta}_L^{(s-1)}+2(s-1)(s+4)\right\}}\right]^{-1/2}.
\end{equation} 
The IR-log divergence in this conformally flat situation was first derived in \cite{Giombi:2013yva}, using the spectral $\zeta-$function to evaluate the trace of the heat kernel at coincident points, for the whole family of bulk higher spin gauge fields. This completely agrees with the UV-log divergence of the boundary computation \cite{Tseytlin:2013fca}. 

To begin our discussion, we will first reproduce the holographic derivation based on Eq.(\ref{ConformallyFlatZZZ}) above in a slightly different way by exploiting the WKB-exactness of the diagonal heat kernel instead.  
The heat kernels for totally symmetric transverse-traceless rank-s tensors in hyperbolic space have long been known \cite{Camporesi:1993mz,Camporesi:1994ga,Gopakumar:2011qs}. A remarkable feature, referred to as WKB exactness, is that after factorization of the exponential of a multiple of the Ricci scalar \footnote{We set the radius of the hyperbolic space to 1, so that the bulk Ricci scalar is simply $\hat{R}=-42$.}, the trace only has finite many terms so that the combined physical (rank s) and ghost (rank s -1) contributions are given by the following proper-time integrals (in the standard route to the \textit{log-dets}.)

\begin{align}
&\int_{0}^{\infty}\frac{dt}{t}\left\{\mbox{tr}_{_{\bot\top}}\,e^{-\left\{\hat{\Delta}_L^{(s)}+2(s-1)(s+4)\right\}\,t}\,-\,\mbox{tr}_{_{\bot\top}}\,e^{-\left\{\hat{\Delta}_L^{(s-1)}+2(s-1)(s+4)\right\}\,t}\,\right\}
\\\nonumber
\\\nonumber
\sim&\int_{0}^{\infty}\frac{dt}{t^{9/2}}\left\{e^{-(s+1)^2t} g(s)\left[1+\frac{2t}{5} \left(s^2+4s+5\right) + \frac{4t^2}{15}(s+2)^2 \right]\right.\\
& \kern 4em - \left. e^{-(s+2)^2t} g(s-1) \left[1+\frac{2t}{5} \left(s^2+2s+2\right) + \frac{4t^2}{15}(s+1)^2 \right]\right\}~, \nonumber
\end{align}\\
where
\begin{equation}
g(s)=\displaystyle{\frac{(s+1)(s+2)^2(s+3)}{12}}
\end{equation}
denotes the number of components of the transverse-traceless field of spin $s$. 

After evaluation of the proper-time integrals in terms of Gamma functions, we obtain the numerical factor corresponding to the one-loop effective Lagrangian on hyperbolic space that accompanies the volume, following the prescription in \cite{Imbimbo:1999bj}. The volume anomaly is given by the Q-curvature so that we can directly read off the type-A Weyl anomaly coefficient, it coincides with the result obtained from the boundary computation on the round six-sphere, written in terms of the number of dynamical degrees of the rank-s CHS in 6D $\nu_s=\frac{(s+1)^2(s+2)^2}{4}$, 

\begin{equation}
a\,=  {\frac{\nu_s}{360\cdot7!}} \left( 88\,{\nu}_s^{3/2}-110\,\nu_s-4\,{\nu}_s^{1/2}+1\right)~. 
\end{equation}

\section{Ricci flat 6D boundary}
In this section we explore deviations from conformal flatness. To proceed with, we first consider Ricci-flat boundary metrics and we assume that the factorization of the higher-derivative kinetic operator into Laplacians still holds with a Lichnerowicz-type coupling to the Weyl tensor, thus the one-loop boundary partition function becomes 
\begin{equation}
Z^{^{1-loop}}_{_{bndry}}=\left[\frac{\left(\det_{\top}\left\{{\Delta}_L^{(s)}\right\}\right)^{s+1}}{\left(\det_{\top}\left\{{\Delta}_L^{(s-1)} \right\}\right)^{s+2}}\right]^{-1/2}
\end{equation}
In order to work out the UV log-divergence, the relevant heat coefficient can be read off from Eq.(B13) in \cite{Liu:2017ruz} for the $(0,s,0)$ representation, where the only curvature invariants that come into play due to Ricci-flatness are $A_5$, $A_{16}$ and $A_{17}$. Alternatively, the very same coefficient can be obtained by our explicit results for the totally symmetric rank-s tensor when the generic dimension $n$ is set to 6 and the trace components are subtracted, that is
\begin{equation}    b_{6,s}^{\top}\Big|_{\text{Ricci-flat}}=\left(b_{6,s}-b_{6,s-2}\right)\Big|_{\text{Ricci-flat}}
\end{equation}
and 
\begin{equation}    b_{6,s-1}^{\top}\Big|_{\text{Ricci-flat}}=\left(b_{6,s-1}-b_{6,s-3}\right)\Big|_{\text{Ricci-flat}}
\end{equation}
with
\begin{align}
\label{B6RicciFlatUnconst}
    7!b_6,s\Big|_{\text{Ricci-flat}}&=  A_5 \left\{-3\cdot\binom{s+5}{5}+56\cdot\binom{s+6}{7}-1260\cdot\binom{s+7}{9}\right\} \\ \nonumber \\
    &\kern-5.5em + A_{16} \left\{\frac{44}{9}\cdot\binom{s+5}{5}-84\cdot\binom{s+6}{7}+420\cdot\binom{s+7}{9}-23520\cdot\binom{s+8}{11}\right\}\nonumber\\ \nonumber \\
    &\kern-5.5em + A_{17} \left\{\frac{80}{9}\cdot\binom{s+5}{5}-168\cdot\binom{s+6}{7}+6720\cdot\binom{s+7}{9}+6720\cdot\binom{s+8}{11}\right\}~.\nonumber
\end{align}
The accumulated heat coefficients for the physical and ghost fields are simply given by  $(s+1)\cdot b_{6,s}^{\top}$ and $(s+2)\cdot b_{6,s-1}^{\top}$, respectively, with 
\begin{align}
    b_{6,s}^{\top}\Big|_{\text{Ricci-flat}}&=  \, g(s) {\frac { 45\,{s}^{4}+360\,{s}^{3}+383\,{s}^{2}-1348\,s+180 }{7!\;60}}\,A_5
 \\ \nonumber \\
    &\kern-3.5em - g(s) {\frac { 28\,{s}^{6}+336\,{s}^{5}+823\,{s}^{4}-2376\,{s}^{3}-5927\,{s}^{2}+9636
\,s-880 }{7!\;180}}  A_{16} \nonumber\\ \nonumber \\
    &\kern-3.5em + 2 g(s) {\frac {{s}^{6}+12\,{s}^{5}+121\,{s}^{4}+648\,{s}^{3}+652\,{s}^{2}-2064\,s+200
 }{7!\;45}}  A_{17} .\nonumber
\end{align}
In all, the UV log-divergence is essentially given by 
\begin{align}
    (s+1)b_{6,s}^{\top}-(s+2)b_{6,s-1}^{\top}\Big|_{\text{Ricci-flat}}&= -\frac {\nu_{s}}{9\cdot 7!} \left(63\,\nu_s -308\,\nu^{1/2}_s + 272\right)A_5
 \\ \nonumber \\
    &\kern-5.5em - \frac {\nu_{s}}{45\cdot 7!}\left( 168\,{\nu_s}^{3/2}-1785\,\nu_s+5376\,{\nu}_s^{1/2}-3979\right) A_{16} \nonumber\\ \nonumber \\
    &\kern-5.5em + \frac {\nu_{s}}{45\cdot 7!}\left( 48\,{\nu_s}^{3/2}+1200\,\nu_s-6264\,{\nu}_s^{1/2}+5416 \right)A_{17} .\nonumber
\end{align}
Now, the structure of the UV-log divergences of any 6D Weyl invariant action is dictated by the trace (or Weyl or conformal) anomaly (see e.g.~\cite{Bastianelli:2000hi}):
\begin{equation}
{\mathcal A}\,=\,-{a}\,E_6\,+\,{c_1}\,I_1\,+\,{c_2}\,I_2\,+\,{c_3}\,I_3
\end{equation}
The restriction to a Ricci-flat, but not conformally flat, background forces two linear relations between the four terms of the anomaly basis, namely, $E_6=64I_1+32I_2$ and $I_3=4I_1-I_2$, so that there are only two independent terms in the anomaly, say
\begin{equation}
{\mathcal A}\,=\,[-64a+c_1+4c_3]\,I_1\,+\,[-32a+c_2-c_3]\,I_2.
\end{equation}
The dictionary is simple in the boundary Ricci-flat case (see e.g. table 1 of reference \cite{Aros_2019JHEP}) $A_5=-I_3=I_2-4I_1$, $A_{16}=I_2$ and $A_{17}=-I_1$, therefore we obtain for the UV log-divergence of the CHS
\begin{eqnarray}
{\mathcal A} &=&  -\frac{\nu_s}{45\cdot 7!} \left( 48\,{\nu}_s^{3/2} -60\,\nu_s -104\,{\nu}_s^{1/2} + 24 \right) I_1 \\
&-&
\frac{\nu_s}{45\cdot 7!} \left( 168\,{\nu_s}^{3/2}-1470\,\nu_s +3836\,{\nu}^{1/2}_s-2619 \right) I_2
 \nonumber
\end{eqnarray} 
Finally, the new information on the central charges amounts to
\begin{eqnarray}
c_1+4c_3 &=&\frac{\nu_s}{45\cdot 7!}\left(656\,{\nu}_s^{3/2} -820\,\nu_s + 72\,{\nu}^{1/2}_s+32\right)
\end{eqnarray} 
and
\begin{eqnarray} 
\qquad\quad c_2-c_3 &=& \frac{\nu_s}{45\cdot 7!}\left( 184\,{\nu}_s^{3/2}+1030\,\nu_s-3852\,{\nu}_s^{1/2}+2623 \right). 
\end{eqnarray} 
As it can be appreciated, the information on the type-B Weyl anomaly coefficients $c_1,\,c_2$ and $c_3$ is partial; essentially $c_3$ remains undetermined. It is worth noticing that $c_3$ is proportional to the $C_T$ coefficient of the two-point function of the stress-energy tensor, $C_T=3024\,c_{3}$. Alternatively, in the convention of \cite{Liu:2017ruz}  $c=\frac{c_2-c_3}{32}$,
$c^{\prime\prime}=\frac{c_1-2c_2+6c_3}{192}=\frac{c_1+4c_3}{192}-\frac{c_2-c_3}{96}$,$c^{\prime}=\frac{c_1-4c_2}{192}=\frac{c_1+4c_3}{192}-\frac{c_2-c_3}{48}-\frac{c_3}{24}$  it is clear that $c$ and $c^{\prime\prime}$ have been fully determined, whereas $c^\prime$ remains unknown on Ricci-flat spaces.

\section{Bulk Poincar\'e-Einstein}
Now we will consider the holographic counterpart, consisting in a Poincar\'e-Einstein metric with a Ricci-flat representative in the family of conformal boundary metrics. The bulk partition function contains the physical and the ghost functional determinants  

\begin{equation}
Z^{^{1-loop}}_{_{bulk}}=\left[\frac{\det_{\bot\top}\left\{\hat{\Delta}_L^{(s)}+2(s-1)(s+4)\right\}}
{\det_{\bot\top}\left\{\hat{\Delta}_L^{(s-1)}+2(s-1)(s+4)\right\}}\right]^{-1/2}
\end{equation}
To compute holographically the type-B central charges, it suffices to consider the Weyl content of the $\hat{b}_6$ anomaly coefficient when restricting to boundary Ricci-flat metrics. The Weyl content of the heat coefficient of the Lichnerowicz Laplacian can be read off from the contribution of the three curvature invariants $\hat{A}_5, \hat{A}_{16}$ and $\hat{A}_{17}$. As already pointed out, they simply follow from extrapolation of the 6D results when written in terms of combinatorial coefficients
\begin{align}
\label{B6BulkRicciFlatUnconst}
     7!\cdot\hat{b}_6,s\big|_{\text{Ricci flat}}&=  \hat{A}_5 \left\{-3\cdot\binom{s+6}{6}+56\cdot\binom{s+7}{8}-1260\cdot\binom{s+8}{10}\right\} \\ \nonumber \\
    &\kern-5.5em + \hat{A}_{16} \left\{\frac{44}{9}\cdot\binom{s+6}{6}-84\cdot\binom{s+7}{8}+420\cdot\binom{s+8}{10}-23520\cdot\binom{s+9}{12}\right\}\nonumber\\ \nonumber \\
    &\kern-5.5em + \hat{A}_{17} \left\{\frac{80}{9}\cdot\binom{s+6}{6}-168\cdot\binom{s+7}{8}+6720\cdot\binom{s+8}{10}+6720\cdot\binom{s+9}{12}\right\}~.\nonumber
\end{align}
Assuming now that the WKB-exactness in the conformally flat case extends to the Weyl terms in the present Poincar\'e-Einstein, we end up with the proper time integrals  
\begin{align}
&\int_{0}^{\infty}\frac{dt}{t}\left\{\mbox{tr}_{_{\bot\top}}\,e^{-\left\{\hat{\Delta}_L^{(s)}+2(s-1)(s+4)\right\}\,t}\,-\,\mbox{tr}_{_{\bot\top}}\,e^{-\left\{\hat{\Delta}_L^{(s-1)}+2(s-1)(s+4)\right\}\,t}\,\right\}
\\\nonumber
\\\nonumber
\sim&\int_{0}^{\infty}\frac{dt}{t^{9/2}}\left\{e^{-(s+1)^2t} \left[t^3 \hat{b}_{6,s}^{\bot\top} + \ldots  \right]- e^{-(s+2)^2t}\left[t^3 \hat{b}_{6,s-1}^{\bot\top} + \ldots  \right]\right\}, \nonumber
\end{align}
By making use of the holographic prescription (see e.g. table 2 of reference \cite{Aros_2019JHEP}), we find the contribution to the boundary Weyl anomaly in terms of the two independent boundary curvature invariants, say $I_1$ and $I_2$, that remain. Let us explain in details how this is done. First, one has to notice that the bulk invariant $\hat{A}_{12}=-42\hat{W}^2$ descends to a multiple of the boundary $R\,W^2$ that vanishes on the Ricci-flat boundary; this restricts the basis of bulk invariants since now the combination corresponding to the above $-4\hat{W}^{'\,3}+\hat{W}^{3}-\hat{\Phi}_7$ does not produce any contribution to the holographic Weyl anomaly. By focusing only on the cubic Weyl terms, for which the holographic prescription is trivially $\hat{W}^{'\,3}\rightarrow I_1$ and $\hat{W}^{3}\rightarrow I_2$, we end up with the same prescription for the hatted quantities as in the boundary, namely $\hat{A}_{17}\rightarrow -I_1$, $\hat{A}_{16}\rightarrow I_2$ and $\hat{A}_{5}\sim\hat{\Phi}_7\sim -4\hat{W}^{'\,3}+\hat{W}^{3} \rightarrow -4I_1+I_2$. 
The answer completely agrees with what we obtained on the boundary under the assumption of factorization of the higher-derivative kinetic operator. Thus we see that they go hand by hand, WKB exactness of the bulk heat kernel and boundary factorization. The obvious reason for the 6D/7D matching can be traced back to the agreement between the heat coefficients $b_{6,s}^{\top}$ and $\hat{b}_{6,s}^{\bot\top}$ when written in terms of the boundary and bulk invariants, respectively,  
when restricted to a Ricci-flat boundary~\footnote{This identity lies at the heart of the functional Schr\"odinger method, see e.g. \cite{Liu:2017ruz}}.

\section{Generic Einstein 6D boundary}
In order to gain access to the remaining $c_3$ coefficient it is necessary to depart from Ricci-flat and conformally flat boundaries. In this section we will explore the natural extension to a generic Einstein 6D boundary and, for practical purposes, we will assume that factorization into Lichnerowicz-type Laplacians still holds on. Therefore, the boundary partition function becomes  \cite{Tseytlin:2013fca}
\begin{equation}\label{Z6DBundary}
Z^{^{1-loop}}_{_{bndry}}=\left[\frac{\displaystyle{\prod_{k'=-1}^{s-1}}\det_{\bot\top}\left\{{\Delta}_L^{(s)}-\frac{R}{30}[s(s+3)+(k'-1)(k'+4)]\right\}}{\displaystyle{\prod_{k=0}^{s-1}} \det_{\bot\top}\left\{{\Delta}_L^{(k)}-\frac{R}{30}[k(k+3)+(s-1)(s+4)]\right\}}\right]^{-1/2}
\end{equation}

When going from a Ricci-flat boundary to a generic Einstein one, in addition to the curvature invariants $A_5$, $A_{16}$ and $A_{17}$, two new ones involving the Weyl tensor come into play, namely $A_{12}$ and $A_{15}$. Since our present interest is the computation of the remaining $c_3$ coefficient, in what follows we will omit the pure-Ricci curvature invariants $A_{10}, A_{11}, A_{13}$ and $A_{14}$ that reduce to multiples of the cubed Ricci scalar $R^3$  and thus only contribute to the $a$ anomaly coefficient.
One possible way to incorporate these two new contributions into $b_6$ is by careful examination of the entire list of traces of the $V's$ coefficients in equation (\ref{b6}), as worked out in \cite{Liu:2017ruz} (see section B.4 therein), for the $(0,s,0)$ representation. 
Remarkably, the overall contributions to $A_{12}$ and $A_{15}$ obtained for $b_{6,s}^{\top}$ in this way exactly match our explicit results for the totally symmetric rank-$s$ tensor in table  \ref{TableOfV} when the generic dimension $n$ is set to 6 and the trace components are subtracted as
\begin{equation}    \left. b_{6,s}^{\top}\right|_{\text{Einstein}}= \left.\left(b_{6,s}-b_{6,s-2}\right) \right|_{\text{Einstein}}
\end{equation}
where $\left.b_{6,s}\right|_{\text{Einstein}}$, now improved by the presence of $A_{12}$ and $A_{15}$, is given by
\begin{align}
\label{B6RicciFlatUnconstGenericEinStein}
   7!\left.b_{6,s}\right|_{\text{Einstein}}&= A_5 \left\{-3\cdot\binom{s+5}{5}+56\cdot\binom{s+6}{7}-1260\cdot\binom{s+7}{9}\right\} \\ \nonumber \\
   & + A_{12} \left\{\frac{14}{3}\cdot\binom{s+5}{5}-98\cdot\binom{s+6}{7}+1680\cdot\binom{s+7}{9}-7560\cdot\binom{s+8}{11}\right\}\nonumber\\ \nonumber \\
   & + A_{15} \left\{-\frac{16}{3}\cdot\binom{s+5}{5} + 476\cdot\binom{s+6}{7} -13440\cdot\binom{s+7}{9} - 60480 \cdot\binom{s+8}{11}\right\}\nonumber\\ & \nonumber \\
    &+ A_{16} \left\{\frac{44}{9}\cdot\binom{s+5}{5}-84\cdot\binom{s+6}{7}+420\cdot\binom{s+7}{9}-23520\cdot\binom{s+8}{11}\right\}\nonumber\\ \nonumber \\
    &+ A_{17} \left\{\frac{80}{9}\cdot\binom{s+5}{5}-168\cdot\binom{s+6}{7}+6720\cdot\binom{s+7}{9}+6720\cdot\binom{s+8}{11}\right\}~.\nonumber
\end{align}
The Weyl content in the heat coefficient for the Lichnerowicz Laplacian on traceless totally symmetric rank-s fields is now given by 
\begin{align}
    b_{6,s}^{\top}&=\, g(s) {\frac {  45\,{s}^{4}+
360\,{s}^{3}+383\,{s}^{2}-1348\,s+180 }{7!\;60}}\,A_5
 \\ \nonumber \\
    &\kern-3.5em - g(s) {\frac { 3\,{s}^{6}+36\,{s}^{5}+33\,{s}^{4}-696\,{s}^{3}-1100\,{s}^{2}+2704\,s
-280
}{7!\;60}}  A_{12} \nonumber
\\ \nonumber \\
    &\kern-3.5em - 2 g(s) {\frac { 3\,{s}^{6}+36\,{s}^{5}+153\,{s}^{4}+264\,{s}^{3}-95\,{s}^{2}-956\,s+40 }{7!\, 45}}  A_{15} \nonumber \\ \nonumber \\
    &\kern-3.5em - g(s) {\frac { 28\,{s}^{6}+336\,{s}^{5}+823\,{s}^{4}-2376\,{s}^{3}-5927\,{s}^{2}+9636
\,s-880 }{7!\;180}}  A_{16} 
\nonumber\\ \nonumber \\
    &\kern-3.5em + 2 g(s) {\frac {{s}^{6}+12\,{s}^{5}+121\,{s}^{4}+648\,{s}^{3}+652\,{s}^{2}-2064\,s+200
 }{7!\;45}}  A_{17} .\nonumber
\end{align}
Notice that the shifts in the Lichnerowicz Laplacians entering the partition function,  Eq.(\ref{Z6DBundary}), modify the accumulated heat coefficient for the physical and ghost fields by terms of the form $R\cdot b^{\top}_{4,s}$. This modifies the overall coefficient of $A_{12}$ as
\begin{equation}
R \cdot b^{\top}_{4,s} = g(s) \frac{9\,{s}^{4}+72\,{
s}^{3}+71\,{s}^{2}-292\,s+56 }{7! \cdot 24}\,A_{12}. 
\end{equation}
Once the shifts are properly taken into account, it only remains to keep track on the dependence on $I_3$ for each curvature invariant. See the extracts of the dictionary in table 1 of reference \cite{Aros_2019JHEP} 
\begin{equation}
A_{5}\sim \frac{5}{3}I_{3},\,  A_{12}\sim -5I_3,\, A_{15}\sim -\frac{5}{6}I_3,\, A_{16}\sim -I_3 \;\mbox{and}\; A_{17}\sim \frac{1}{4}I_3.    
\end{equation}
Since the structure of the trace anomaly on the generic Einstein 6D manifold
gets rearranged in a pure-Ricci term, that is $R^3$ which is essentially the 6D Q-curvature, and shifted coefficients for the three Weyl invariants  
\begin{equation}
{\mathcal A}\,=\,-16{a}\,R^3\,+\,{(c_1-96\,a)}\,I_1\,+\,{(c_2-24\,a)}\,I_2\,+\,{(c_3+8\,a)}\,I_3,
\end{equation}
the overall coefficient of $I_3$ grants access to the sought-after $c_3$ charge 

\begin{equation}\label{c3-Amended}
c_3 = -\frac{\nu_s}{45\cdot 7!}\left(76\,{\nu}_s^{3/2}+745\,\nu_s - 3548\,{\nu}^{1/2}_s + 2637\right).
\end{equation}

\section{Bulk Poincar\'e-Einstein with Einstein boundary} 

For the corresponding space-filling metric, whose boundary has been discussed in the previous section, we choose the Poincar\'e-Einstein metric
\begin{equation}
    \hat{g}_{_{\text{PE/E}}} = \displaystyle{\frac{dx^2+(1-\lambda x^2)^2 g_{_{\text{E}}}}{x^2}}
\end{equation}
where $x=0$ defines the boundary,  $g_{_{\text{E}}}$ is the Einstein metric on the conformal boundary and $\lambda =R/120$ is a multiple of the (necessarily constant) Ricci scalar $R$ of the (6-dimensional) boundary Einstein metric. With that in mind, we examine the bulk partition function given by  \cite{Tseytlin:2013fca,Giombi:2013yva}
\begin{equation}
Z^{^{\text{1-loop}}}_{_{\text{bulk}}}=\left[\frac{\det_{\bot\top}\left\{\hat{\Delta}_L^{(s)}+2(s-1)(s+4)\right\}}
{\det_{\bot\top}\left\{\hat{\Delta}_L^{(s-1)}+2(s-1)(s+4)\right\}}\right]^{-1/2}
\end{equation}

It is worth noticing that, in correspondence with the boundary computation, there are again two additional bulk curvature invariants $\hat{A}_{12}$ and $\hat{A}_{15}$ that come into play. Thus, for the unconstrained bulk field we obtain 
\begin{align}
\label{b7d}
   7!\hat{b}_{6,s}&= \hat{A}_5 \left\{-3\cdot\binom{s+6}{6}+56\cdot\binom{s+7}{8}-1260\cdot\binom{s+8}{10}\right\} \\ \nonumber \\
   & + \hat{A}_{12} \left\{\frac{14}{3}\cdot\binom{s+6}{6}-98\cdot\binom{s+7}{8}+1680\cdot\binom{s+8}{10}-7560\cdot\binom{s+9}{12}\right\}\nonumber\\ \nonumber \\
   & + \hat{A}_{15} \left\{-\frac{16}{3}\cdot\binom{s+6}{6} + 476\cdot\binom{s+7}{8} -13440\cdot\binom{s+8}{10} - 60480 \cdot\binom{s+9}{12}\right\}\nonumber\\ & \nonumber \\
    &+ \hat{A}_{16} \left\{\frac{44}{9}\cdot\binom{s+6}{6}-84\cdot\binom{s+7}{8}+420\cdot\binom{s+8}{10}-23520\cdot\binom{s+9}{12}\right\}\nonumber\\ \nonumber \\
    &+ \hat{A}_{17} \left\{\frac{80}{9}\cdot\binom{s+6}{6}-168\cdot\binom{s+7}{8}+6720\cdot\binom{s+8}{10}+6720\cdot\binom{s+9}{12}\right\}~.\nonumber
\end{align}

Furthermore, there is an additional contribution to $\hat{A}_{12}$ coming from 
\begin{equation}
    \hat{R}\cdot\hat{b}_{4,s} = \hat{A}_{12} \left\{\frac{1}{180} \left(\begin{array}{c}
                               s+6 \\
                               6
                             \end{array}
\right)-\frac{1}{12} \left(\begin{array}{c}
                               s+7 \\
                               8
                             \end{array}
\right)+
\frac{3}{2}  \left(\begin{array}{c}
                               s+8 \\
                               10
                             \end{array}
\right)\right\}
\end{equation}

To determine the Weyl content of the IR-log divergence, under the assumption of WKB-exactness of the heat kernel for transverse traceless bulk fields, we end up with the proper-time integrals  
\begin{align}
&\int_{0}^{\infty}\frac{dt}{t}\left\{\mbox{tr}_{_{\bot\top}}\,e^{-\left\{\hat{\Delta}_L^{(s)}+2(s-1)(s+4)\right\}\,t}\,-\,\mbox{tr}_{_{\bot\top}}\,e^{-\left\{\hat{\Delta}_L^{(s-1)}+2(s-1)(s+4)\right\}\,t}\,\right\}
\\\nonumber
\\\nonumber
\sim&\int_{0}^{\infty}\frac{dt}{t^{9/2}}\left\{e^{-(s+1)^2t} \left[t^2 \hat{b}_{4,s}^{\bot\top} + t^3 \hat{b}_{6,s}^{\bot\top} + \ldots  \right]- e^{-(s+2)^2t}\left[t^2 \hat{b}_{4,s-1}^{\bot\top} + t^3 \hat{b}_{6,s-1}^{\bot\top} + \ldots  \right]\right\}. \nonumber
\end{align}
It suffices to keep track on the dependence of the bulk curvature invariants on the Fefferman-Graham invariant $\hat{\Phi}_7$ if we are only interested in the $c_3$ coefficient of the holographic anomaly. Again, from the dictionary in table 2 of reference \cite{Aros_2019JHEP}, the contribution to $\hat{\Phi}_7$ from every bulk curvature invariant is pinned down  
\begin{equation}
\hat{A}_{5}\sim 3\hat{\Phi}_{7},\,  \hat{A}_{12}\sim -\frac{21}{2}\hat{\Phi}_{7},\, \hat{A}_{15}\sim -\frac{3}{2}\hat{\Phi}_{7},\, \hat{A}_{16}\sim -\frac{3}{2}\hat{\Phi}_{7},\;\text{ and }\; \hat{A}_{17}\sim \frac{3}{8}\hat{\Phi}_{7}.
\end{equation}

The prescription to compute the holographic Weyl anomaly maps the bulk $\hat{\Phi}_7$ to the boundary $\Phi_6$. Additionally, on the Einstein boundary  $I_3=16I_1-4I_2+3\Phi_6$ is satisfied and therefore the overall coefficient of $\hat{\Phi}_7$ is simply three times that of $I_3$, \textit{i.e.}, $3(c_3+8\,a)$.

Finally, the results above then yield a holographically derived $\tilde{c}_3$ given by
\begin{equation}\label{c3-bulk}
\tilde{c}_3 = -\frac{\nu_s}{45\cdot 7!}\left(148\,{\nu}_s^{3/2}+25\,\nu_s - 1604\,{\nu}^{1/2}_s + 1341\right), 
\end{equation}
which differs from its boundary counterpart in Eq.(\ref{c3-Amended}). The difference is given by
\begin{equation}
c_3-\tilde{c}_3 = \frac{8\,\nu_s}{5\cdot 7!}\left({\nu}_s^{1/2}-1\right)\left({\nu}_s^{1/2}-3\right)\left({\nu}_s^{1/2}-6\right). \label{mismatch}
\end{equation}
It must be noticed that the zeros of the expression above correspond to $s=0,1$ and $2$, as expected, since for these values the mapping has already been verified. The enforced agreement for these values of the spin essentially fixes the polynomial up to an overall factor. For comparison purposes, it would then be desirable to explore an alternative, third, route~\footnote{One concrete possibility relies on the computation of R\'enyi entropy coefficients~\cite{Beccaria_2017}.} to $c_3$ or, equivalently, to $C_T$.

\section{Conclusion and prospects}
In this work we continued the study of conformal higher spin gauge fields and their holographic dual given by massless higher spin fields in the bulk. We considered the boundary to be an Einstein manifold, and therefore Bach-flat, and a dual Poincar\'e-Einstein bulk metric. It was also assumed a minimal departure from the conformally flat setting by allowing a Lichnerowicz-type coupling with both boundary and bulk Weyl tensors.  

For totally symmetric rank-$s$ tensors we computed the $b_{6,s}$ heat coefficient for the Lichnerowicz Laplacian on a generic Einstein background in any dimension. With this key ingredient, we were able to probe the holographic formula for one-loop partition functions under the assumptions of factorization of the CHS higher-derivative kinetic operator and WKB-exactness of the heat kernel for the MHS bulk field. For a  Ricci-flat boundary, we were able to show the exact matching between the UV-log and IR-log divergences of the boundary and bulk partition functions, respectively. In addition, we were also able to obtain partial information on the type-B anomaly coefficients: determined $c_1$ and $c_2$ in terms of $c_3$, but left the latter undetermined. It would be interesting to extend these results to include half-integer spins and to check whether the combinatorial structure of the traces in $b_6$, obtained by the group theory method of \cite{Liu:2017ruz}, admits a ready extrapolation to generic dimensions.

Considering a generic Einstein boundary, beyond Ricci-flatness, grants access to $c_3$. In this case, however, we found a mismatch between the boundary and bulk computations. This mismatch does not affect the functional dependence of $c_1$ and $c_2$ on $c_3$. 
We must point out that the validity of our results on the boundary relies on the factorization conjecture of the CHS kinetic operator, which has been questioned \cite{Nutma:2014pua}. In particular, it has been argued that the factorization is obstructed at first order in the background curvature. 

However, as in the 4D case, whether the obstruction terms modify the type-B Weyl anomaly coefficients $c's$ remains an open issue. Moreover, it must be stressed that the argument in \cite{Nutma:2014pua} was under the assumption that the CHS kinetic operator is diagonal and gauge invariant for each $s$ independently, which seems to fail in general beyond the small curvature expansion~\cite{Grigoriev:2016bzl}.  For instance, in \cite{Beccaria:2018rxp} it was shown that for the spin $3$ kinetic term, the gauge invariance can be attained by switching on a coupling to a conformal spin$-1$ field. This produces a shift in the anomaly $c$-coefficients. The vanishing of the mismatch on Ricci-flat backgrounds might be attributed to the fact that terms obstructing the factorization have no bearing on the log-divergence, and hence, on the Weyl anomaly.     
 
 In all, even though the program of constructing CHS gauge fields in a curve background is far from being settled, even for the spin-3, we expect that once the mixing issue is resolved a proper holographic picture will come up for free fields and our present results, although somewhat preliminary, will prove useful in that endeavour. Let us close by mentioning an interesting approach to bypass the problem of factorization and further study the mixing of CHS that has been recently proposed in \cite{Kuzenko:2020opc}. This shifts the focus to maximal depth CHS fields and decreases the order of the kinetic operator. The bulk dual clearly corresponds to a (partially-)massive higher-spin field, but the details of the holographic picture must yet be unraveled. Constraints on the gravitational background are to be expected according to the analysis in \cite{Cortese:2013lda} (or, more recently, \cite{Rahman:2020qal}), but then again this analysis overlooked the possibility of a mixing with lower spins in order to maintain Weyl invariance. It would be interesting to investigate these alternatives in the future.    
 
 %%%%%%%%%%%%%%%%%%%%%%%%%%%%%%%%%%%%%%%%%%%%%%%%%%%%%%%%%%%%%%

\ack
We would like to thank P. Sundell for useful conversations.
\appendix

\section{Traces} \label{AppendixTraces}

Here we collect the curvature invariants that contribute to the Weyl content of the $b_6$ heat coefficient for $-\nabla^2-E$ when acting on unconstrained tensor fields, as given by \cite{Gilkey:1975iq,Bastianelli:2000hi}. When restricted to an Einstein metric, the only contributions come from the invariants $A_5, A_{12}, A_{15}, A_{16}$ and $A_{17}$ ($A_9=-A_5$ modulo a trivial total derivative, the same holds for the traces $V_3=-V_1$ and $V_{10}=-V_9$) 

\begin{align}\label{b6}
{b}_6\{-{\nabla}^{2}-E\}\,=&\, \frac{1}{(4\pi)^3 7!}\,\mbox{tr}_V \bigg\{ -3A_5 + \frac{14}{3}A_{12} - \frac{16}{3}A_{15} + \frac{44}{9}A_{16} + \frac{80}{9}A_{17}
\\\nonumber\\
\nonumber &+ 14\bigg( -4 V_1 - 12V_4 + 6V_5 - 4V_6 + 5V_7  \\\nonumber\\
\nonumber & - 30 V_{10}  + 60V_{11} + 30V_{12} + 30V_{16} + 2V_{20}\bigg)\bigg\}
\end{align}

\textbf{Endomorphism} The index structure of the endomorphism (Lichnerowicz coupling containing the Weyl tensor) is as follows (see e.g. \cite{Tseytlin:2013jya,Cortese:2013lda})
\begin{equation}
E= s(s-1) R^{\rho_{1}\,\rho_{2}}_{~\nu_{1}~\nu_{2}}\delta_{\rho_{1}}^{(\mu_{1}}\delta_{\rho_{2}}^{\mu_{2}}\delta_{\nu_{3}}^{\mu_{3}}\cdots\delta_{\nu_{s}}^{\mu_{s})} - s R^{\rho_{1}}_{~\nu_{1}}\delta_{\rho_{1}}^{(\mu_{1}}\delta_{\nu_{2}}^{\mu_{2}}\delta_{\nu_{3}}^{\mu_{3}}\cdots\delta_{\nu_{s}}^{\mu_{s})} \\
\end{equation}

\textbf{Connection curvature} The curvature of the spin connection for the rank$-s$ totally symmetric tensor is given by (see e.g.\cite{Tseytlin:2013jya}) 
\begin{equation}
F= \frac{1}{2}R_{\mu\nu}^{\;\;\; ab}\,\Sigma_{ab}
\end{equation}
with $\Sigma_{ab}$ realizing the Lorentz algebra
\begin{align}
[\Sigma_{ab} ,\Sigma_{cd}]=-\eta_{ac}\Sigma_{bd}+\eta_{ad}\Sigma_{bc}+\eta_{bc}\Sigma_{ad}-\eta_{bd}\Sigma_{ac}
\end{align}\\
The explicit index structure of the curvature \footnote{This corrects a misprint in equation (A17) in \cite{Acevedo:2017vkk}.}, generalizing spins one and two, is given by
\begin{equation}
(\Sigma_{ab})_{c_{1}\cdots c_{s}}^{d_{1}\cdots d_{s}}= 2s\, \delta_{[a}^{(d_{1}}\eta_{b](c_{1}}\delta_{c_{2}}^{d_{2}}\cdots\delta_{c_{s})}^{d_{s})}
\end{equation}

\setlength\extrarowheight{10pt}
\begin{table}[h]
\begin{center}
\begin{tabular}{lll}\hline
$\mbox{tr}_V V_{1}$ & $=
    \mbox{tr}_V |{\nabla}{F}|^{2}$ & $= \displaystyle{-\binom{n+s}{n+1}\cdot A_5}$\\\hline
$\mbox{tr}_V V_{4}$ & $= 
    \mbox{tr}_V {F}^{3}$ & $= \phantom{-}\displaystyle{\binom{n+s}{n+1}\cdot A_{17}}$\\\hline 
$\mbox{tr}_V V_{5}$ & 
    $= \mbox{tr}_V \text{Riem }{F}^{2}$ & $= \displaystyle{-\binom{n+s}{n+1}\cdot A_{16}}$\\\hline
$\mbox{tr}_V V_{6}$ & 
    $= \mbox{tr}_V \text{Ric }{F}^{2}$ & $= \displaystyle{-\binom{n+s}{n+1}\cdot A_{15}}$\\\hline
$\mbox{tr}_V V_{7}$ & 
    $= \mbox{tr}_V \text{R }{F}^{2}$ & $= \displaystyle{-\binom{n+s}{n+1}\cdot A_{12}}$\\\hline 
$\mbox{tr}_V V_{10}$ & $= \mbox{tr}_V |{\nabla}{E}|^{2}$ & $=
    \displaystyle{3\binom{n+1+s}{n+3}\cdot A_5}$\\\hline 
\multirow{4}{4em}{$\mbox{tr}_V V_{11}$} & 
    \multirow{4}{4em}{$=\mbox{tr}_V{E}^{3}$} &
    $=\phantom{-}\displaystyle{\left[8\binom{n+2+s}{n+5}+8\binom{n+1+s}{n+3}\right]\cdot A_{17}}$\\ 
    & & 
    $\phantom{=}\displaystyle{-\left[28\binom{n+2+s}{n+5}+\binom{n+1+s}{n+3}\right]\cdot A_{16}}$\\ 
    & & 
    $\phantom{=}\displaystyle{-\left[72\binom{n+2+s}{n+5}+18\binom{n+1+s}{n+3}\right]\cdot A_{15}}$\\
    & &
    $\phantom{=}\displaystyle{-9\binom{n+2+s}{n+5}\cdot A_{12}}$\\\hline
\multirow{2}{5em}{$\mbox{tr}_V V_{12}$} & 
    \multirow{2}{5em}{$= \mbox{tr}_V {E}{F}^{2}$} & 
    $= \displaystyle{3\binom{n+1+s}{n+3}\cdot A_{16}}
    \displaystyle{+\left[4\binom{n+1+s}{n+3}+\binom{n+s}{n+1}\right]\cdot A_{15}}$\\ 
    & & 
    $\phantom{=}\displaystyle{+ \binom{n+1+s}{n+3}\cdot A_{12}}$\\\hline 
$\mbox{tr}_V V_{16}$  & 
    $= \mbox{tr}_V \text{R }{E}^{2}$ & 
    $= \displaystyle{3\binom{n+1+s}{n+3}\cdot A_{12}}$\\\hline
$\mbox{tr}_V V_{20}$ & $= \mbox{tr}_V \text{Riem}^{2}{E}$ &
    $= \displaystyle{-\binom{n+s}{n+1}\cdot A_{12}}$\\\hline
\end{tabular}
 
\caption{Traces by explicit contraction for unconstrained totally symmetric rank-$s$ tensors}\label{TableOfV} 
\end{center}
\end{table}
Most of the traces reduce to the quadratic ones already computed in a previous paper \cite{Acevedo:2017vkk}. The challenging ones are $V_4, V_{11}$ and $V_{12}$. We first contracted the permanents\footnote{The relevant identities are Eqs.(A11-A13) in \cite{Acevedo:2017vkk}, but for cubic contractions of the Weyl tensor one must subtract traces in the spectator indices in Eq.(A11).} and then used the Cadabra code~\cite{Peeters:2006kp,Peeters:2007wn} to determine the final contractions and read off the coefficients of the curvature invariants in Table~\ref{TableOfV}.  

\section{Heat coefficient from group theory method}

In 6D, the traces entering the heat coefficients were computed via so-called group theory method~\cite{Liu:2017ruz}. To make contact with our current computation, one must choose the representation with Dynkin labels $(a, b, c)=(0,s,0)$ that corresponds to the totally symmetric traceless rank-s tensor. 

In general, the dimension of the representation is given by
\begin{equation}
\textrm{Dim}_R(a, b, c) = \frac{1}{12} (a + 1)(b + 1)(c + 1)(a + b + 2)(b + c + 2)(a + b + c + 3).
\end{equation}
Using the Weyl character formula it can be shown that the second order Dynkin index is given by
\begin{equation}
I_2(a, b, c) = \frac{\textrm{Dim}_R}{60} (3a^2 + 2a(2b + c + 6) + 4b^2 + 4b(c + 4) + 3c(c + 4)). 
\end{equation}
Analogously, the third and fourth order generalizations are given by \cite{Okubo:1981td} 
\begin{equation}I_3(a,b,c)=\frac{\textrm{Dim}_R}{120}(a-c)(a+c+2)(a+2b+c+4)\end{equation}
\begin{eqnarray}
I_4(a,b,c)&=&\frac{\textrm{Dim}_R}{3360}(3a^4 + 8a^3b + 4a^3c + 24a^3 + 2a^2b^2\nonumber \\
&&  + 2a^2bc + 30a^2b-4a^2c^2 + 6a^2c + 54a^2-2ab^3-18ab^2c\nonumber\\
&& -50ab^2+2abc^2-28abc-34ab+4ac^3+6ac^2-2ac+24a\\
&& -6b^4-12b^3c-48b^3+2b^2c^2-50b^2c-122b^2+8bc^3+30bc^2\nonumber\\
&&-34bc-104b+3c^4+24c^3+54c^2+24c)\nonumber
\end{eqnarray}

These results determine $b_{6,s}^{\top}$ for Ricci-flat geometries, given by
\begin{align}
\label{B6LiuMcPeak}
    7!\cdot b_6(R)\Big|_{R_{ab} = 0}\,=\, 
     & A_5 \left(\frac{3150
   I_2^2}{17\text{Dim}_R}+9\text{Dim}_R-\frac{1827 I_2}{17}+\frac{700 I_4}{17}\right)
    \\
    &\kern-7.5em + A_{16} \left(-\frac{9150 I_2^3}{17\text{Dim}_R^2}+\frac{12900
   I_2^2}{17 \text{Dim}_R}-\frac{560 I_2 I_4}{17 \text{Dim}_R}-\frac{1010
   I_3^2}{\text{Dim}_R}+\frac{44\text{Dim}_R}{9}-\frac{8597 I_2}{51}+\frac{14140
   I_4}{51}\right)\nonumber\\
   &\kern-7.5em+A_{17}\left(-\frac{1725 I_2^3}{17\text{Dim}_R^2}+\frac{10875 I_2^2}{17
   \text{Dim}_R}-\frac{10640 I_2 I_4}{17\text{Dim}_R}+\frac{2185 I_3^2}{\text{Dim}_R}+\frac{80\text{Dim}_R}{9}-\frac{17804 I_2}{51}-\frac{2240 I_4}{51}\right).\nonumber
\end{align}

\section*{References}
\bibliographystyle{JHEP.bst}
\bibliography{bibTe.bib}

\end{document}